\documentstyle[12pt]{article}
\begin{document}
\thispagestyle{empty}
\begin{flushright}
Imperial/TP/94-95/13
\end{flushright}
\begin{center}
\vspace{3.0cm}
{\Large\bf QCD running coupling constant \\
\vspace{0.3cm}
in the timelike region } \\
\vspace{2.0cm}
{\large\bf     H.F. Jones} \\
\vspace{0.2cm}
{\em Physics Department, Imperial College, London SW7 2BZ,\\
United Kingdom }

\vspace{0.2cm}
{\large\bf I.L. Solovtsov   }\\
\vspace{0.2cm}
{\em  Bogoliubov Laboratory  of  Theoretical  Physics, \\
Joint  Institute  for Nuclear Research, \\
Dubna, Moscow region, 141980, Russia } \\
\end{center}
\vspace{2.0cm}
\begin{abstract}
By using a non-perturbative expansion and the dispersion relation
for the Adler $D$--function we propose a new method for constructing the
QCD effective coupling constant in the timelike region.
\end{abstract}

\newpage

The perturbative expansion  is a powerful tool for performing
calculations in quantum chromodynamics. Perturbation theory
with renormalization group improvement has been widely applied to
the description of various processes in both the spacelike and timelike
regions. However, as is well known, the running coupling constant
$\alpha_S (Q^2)$, the expansion parameter in QCD perturbation
theory, is usually defined in the Euclidean range of momentum. If one
wishes to know the ``running coupling constant" in the timelike region,
and how it evolves, one must make use of some analytic
continuation from the Euclidean (~spacelike~) region to the timelike region.
To this end one usually applies the dispersion relation for the Adler
$D$--function [1]
\begin{equation}
\label{Eq.1}
D(q^2)\,=\,-\,q^2\,\int_0^{\infty}\,\frac{d\,s}{{(s-q^2)}^2}\,R(s)\,\, ,
\end{equation}
where
\begin{equation}
\label{Eq.2}
D(q^2)\,=\,q^2\,(-\,\frac{d}{d\,q^2})\,\Pi(q^2)
\end{equation}
and the function $\Pi (q^2)$ is defined by
\begin{equation}
\label{Eq.3}
{\rm i}\,\int d^4x\, \exp ({\rm i}\, q.x)\,
\bigl< 0|\,T\,\{J_{\mu}(x)\,J_{\nu}(0)\}|0\bigr>\,
\propto \,(q_{\mu}\,q_{\nu}\,-\,g_{\mu \nu}\,q^2)\,\Pi(q^2)
\end{equation}
and
\begin{eqnarray}
\label{Eq.4}
R(s)\,&=&\,\frac{1}{2\pi{\rm i}}\,
\bigl[\,\Pi (s+{\rm i}\,\epsilon)\,-\,\Pi (s-{\rm i}\,\epsilon)\,\bigr]\,
=\,\frac{1}{\pi}\,{\rm Im}\, \Pi (s)\nonumber\\
&=&\,-\,\frac{1}{2\pi{\rm i}}\,
\int _{s-{\rm i}\,\epsilon} ^{s+{\rm i}\,\epsilon} \frac{d\,z}{z}\,D(z)\, .
\end{eqnarray}
The assumptions for writing these equations are the following: the
function $\Pi (q^2)$ is an analytic function in the complex $q^2$--plane
with a cut along the positive real axis,
and the dispersion relation for the $D$--function does not require any
subtractions. The contour in Eq.~(4) goes from the point
$z\,=\, s-{\rm i}\,\epsilon$ to the point $z\,=\, s+{\rm i}\,\epsilon$
and lies in the region where the function $D(z)$ is an analytic function
of $z$ (~see Figure~1~).

Let us represent the functions $D(q^2)$ and $R(q^2)$ in the form
\begin{eqnarray}
\label{Eq.5}
D(q^2)\,&\propto &\, \sum_{f}\,Q_f^2\,\bigl[\,1\,+\,d_0\,\lambda_t(q^2)\,+
\,d_1\,\lambda_t^2(q^2)\,+ \cdots \,\bigr] \nonumber\\
\,&\equiv &\,\sum_{f}\,Q_f^2\,\bigl[\,1\,+\,d_0\,{\lambda}_t^{\rm eff}(q^2)\,
\bigr]\,  ,
\end{eqnarray}
and
\begin{eqnarray}
\label{Eq.6}
R(s)\,&\propto& \, \sum_{f}\,Q_f^2\,\bigl[\,1\,+\,r_0\,\lambda_s(s)\,+
\,r_1\,\lambda_s^2(s)\,+\cdots\,\bigr] \nonumber\\
\,&\equiv &\,\sum_{f}\,Q_f^2\,\bigl[\,1\,+\,r_0\,{\lambda}_s^{\rm eff}(s)\,
\bigr]\,  ,
\end{eqnarray}
where we use the definition $\lambda\,=\,\alpha_S/(4\pi)$. The index $t$
means ``$t$--channel" (~the spacelike region~) and $s$ means ``$s$--channel"
(~the timelike region~). The coefficients $d_0$ in Eq.~(5) and $r_0$ in
Eq.~(6) are numerically equal; thus Eq.~(1) leads to a dispersion relation
relating the effective coupling constants in the $t$ and $s$ channels:
\begin{equation}
\label{Eq.7}
\lambda_t^{\rm eff}(q^2)\,=\,-\,q^2\,\int_0^{\infty}\,
\frac{d\,s}{{(s-q^2)}^2}\,\lambda_s^{\rm eff}(s)\,\, .
\end{equation}
The solution of Eq.~(7), given the analytic properties of
$\lambda_t^{\rm eff}(t)$, is
\begin{equation}
\label{Eq.8}
\lambda_s^{\rm eff}(s)\,=\,-\,\frac{1}{2\pi {\rm i}}\,
\int _{s-{\rm i}\,\epsilon} ^{s+{\rm i}\,\epsilon} \frac{d\,z}{z}\,
\lambda_t^{\rm eff}(z)
\end{equation}

Thus, Eq.~(8) serves to define the running coupling
constant in the timelike region. If we use the leading one--loop
approximation for the running coupling constant
$\lambda_t^{\rm eff}$ in the $t$--channel,
\begin{equation}
\label{Eq.9}
\lambda_t^{\rm eff}(z)\,=\,\frac{1}{b_0\,\ln (-z/\Lambda^2),}
\end{equation}
where
$b_0\,=\,11-2n_f/3$ is the first coefficient of the $\beta$--function
and $n_f$ is the number of flavours, we obtain
\begin{equation}
\label{Eq.10}
b_0\,\lambda_s^{\rm eff}(s)\,=\,\frac{1}{\pi}\,{\rm arctan}\,
\frac{\pi}{\ln(s/\Lambda^2)  }\,
=\,\frac{1}{\ln (s/\Lambda^2)} \,-\,\frac{1}{3}\,
\frac{\pi^2}{\ln ^3(s/\Lambda^2) }\,+\cdots
\end{equation}
Thus, one might conclude that the main effect of the analytic
continuation from the spacelike to the timelike region is an
additional contribution (~the so-called $\pi^2$--term~) of
order $1/\ln^3(s/\Lambda^2)$.
The $\pi^2$--terms play an important role in the analysis of various
processes. For example, in the $\tau$--lepton region, where
$\alpha_S\,\simeq\,0.35$ (~see, for example, a discussion in Ref.~[2]~),
the $\pi^2$--correction is about twenty percent and the ratio
$R(s)\,=\,\lambda_s(s)/\lambda_t(q^2)\,\simeq\,0.8$ for
$s\,=\,-\,q^2\,=\,M_{\tau}^2$.
The $\pi^2$--terms  and the problem of the construction of the
running coupling constant in the timelike domain have been discussed
in detail in Refs.~[3-7].

However, the one--loop approximation to $\lambda_t^{\rm eff}$
breaks the analytical properties of $\lambda_t^{\rm eff}(z)$ for which
we wrote the solution (8). Indeed, the function $\lambda_t^{\rm 1-loop}$
has an infrared singularity at $z\,=\,-\,\Lambda^2$, which contradicts
the assumption that the function $\lambda_t^{\rm eff}(z)$ has only a cut
in the complex $z$-plane.
A consequence of this is the fact that Eq.~(7) does not reproduce
the original one--loop coupling constant in the $t$--channel if we use
for $\lambda_s^{\rm eff}(s)$ the ``solution" in the form of Eq.~(8).
In order to maintain the analytic properties of $\lambda^{\rm eff}$
one must make some modification to the initial coupling constant,
which is to be substituted into Eq.~(8).

{}From a phenomenological point of view such modifications can be made in
an {\sl ad hoc} manner, with no fundamental justification. In the present
note, however, we construct the running coupling constant in the timelike
domain in the well-motivated framework of the non-perturbative method
proposed in Ref.~[8], which allows one systematically to determine the
low energy structure in QCD.
The method is based on the introduction of a new small expansion parameter.
For small coupling constant $\alpha_S(Q^2)$ the series thus obtained coincides
with standard perturbation theory, so that high energy physics is essentially
unaffected. Moreover the method continues to be valid in the nonperturbative
region, where $\alpha_S(Q^2)\, \geq \, 1$, since the expansion parameter in
this approach remains small.

The approach is related to the method of the Gaussian effective potential
[9-12], the linear $\delta$--expansion [13,14] and the
approach based on variational perturbation theory [15,16].
Within these approaches the investigated quantity is written from
the very beginning in the form of a series which provides a well-determined
algorithm for
calculating corrections up to any order. In addition, the series contains
some auxiliary parameters which allow one to control the convergence
properties of the series and to construct an expansion based on a new
small parameter, as in this paper. The exact value of a given quantity
is independent of these parameters. However, the corresponding approximate
result contains a residual dependence upon those variables, and we may choose
the latter so as to provide the best approximation to the quantity.
A similar situation holds in standard perturbation theory
for the total cross section for $e^+ \, e^-$ annihilation into hadrons:
the physical
quantity $R_{e^+ e^-}$ is independent of a renormalization scheme,
whereas a finite perturbative approximation to $R_{e^+ e^-}$ is
scheme dependent (~in this connection see Refs.~[17,18]~).

The method described in Ref.~[8] starts with the standard action of QCD,
written as
\begin{equation}
\label{Eq.11}
S(A,q,\varphi)\,=\,S_2(A)\,+\,S_2(q)\,+\,S_2(\varphi)\,+\,
g\,S_3(A,q,\varphi)\,+\,g^2\,S_4(A)\,\, ,
\end{equation}
where $S_2(A)$, including the gauge--fixing term, $S_2(q)$ and $S_2(\varphi)$
are the standard free actions of the gluon, quark and ghost fields, while
$S_3(A,q,\varphi)$ contains the trilinear vertices and $S_4(A)$ the four-point
gluon vertex. By use of an auxiliary field $\chi_{\mu\nu}$ the latter action
can be rewritten as a trilinear action between $A$ and $\chi$, so that $S$
becomes
\begin{equation}
\label{Eq.12}
S(A,q,\varphi,\chi)\,=\,S_2(q)\,+\,S_2(\varphi)\,+\,S_2(\chi)\,
+\,S(A,\chi)\,+\,g\,S_3(A,q,\varphi)\, .
\end{equation}

We then introduce a new split between the free and interaction parts of the
action according to
\begin{equation}
\label{Eq.13}
S(A,q,\varphi,\chi)\,=\,S_0'(A,q,\varphi,\chi)\,+\,S_I'(A,q,\varphi,\chi)\, ,
\end{equation}
where
\begin{equation}
\label{Eq.14}
S_0'\,=\,\zeta^{-1}
\bigl[S(A,\chi)+S_2(q)+S_2(\varphi)\bigr]+ \xi^{-1}\, S_2(\chi)\, ,
\end{equation}
and
\begin{equation}
\label{Eq.15}
S_I'\,=\,g\,S_3(A,q,\varphi)-(\zeta^{-1}-1)\,
\bigl[ \,S(A,\chi)\,+
\,S_2(q)+S_2(\varphi)\bigr]-({\xi}^{-1}-1)\,S_2(\chi)\, ,
\end{equation}
thus introducing two variational parameters $\xi$, $\zeta$. However,
it turns out [8] that the two parameters must be related by
$\xi=\zeta^3$ in order to preserve gauge invariance. After a rescaling of
the fields we obtain the following expression for a general Green function:
\begin{equation}
\label{Eq.16}
G(\cdots)\,=\,\int\, D_{QCD}\,(\cdots)\,\, V(A,q,\varphi)\, \exp(i\,S_0)\,  ,
\end{equation}
with
\begin{eqnarray}
\label{Eq.17}
V(A,q,\varphi)\, &=&\,\sum_n \,\sum_{k=0}^{n} \frac{1}{(n\,-\,k)\,!}
{  \bigl(\,-\,\frac{ \partial }{ \partial \,\kappa } \bigr) }^{n\,-\,k}
\, \frac{i^k}{k!} \,\,
\frac{1}{  [\,1\,+\,\kappa\, (\zeta^{-1}-1)\,]^{\nu/2}}
\nonumber \\
&\times & \,{ \bigl[ g_3 \,S_3(A,q,\varphi ) \bigr]}^k
\exp \bigl\{ \,i\,\bigl[ \,g_4^2\,S_4(A)\,\bigr] \,\bigr\}\, ,
\end{eqnarray}
where
\begin{equation}
\label{Eq.18}
g_3\,=\,\frac{g}{\bigl[\,1\,+\,\kappa\,
({\zeta}^{-1}-1\,)\,\bigr]^{3/2}  }\,\, ,\quad
\quad g_4\,=\,\frac{g}{  {\bigl[\,1\,+\,\kappa\,
({\xi}^{-1}-1)}\bigr]^{1/2}  }
\end{equation}
Here $S_0$ is the standard free action of QCD and $\kappa$ is a parameter
introduced for convenience which is set equal to $1$ at the end of the
calculation.

Analysis of the structure of this variational perturbation series shows
that it can be organized in powers of the new small parameter
$a \equiv 1-\zeta$ if the standard coupling constant $g$ is related to $a$ by
\begin{equation}
\label{Eq.19}
\lambda\,=\,\frac{g^2}{{(4\pi)}^2}\,=\,\frac{1}{C}\,\frac{a^2}{{(1-a)}^3}\, ,
\end{equation}
where $C$ is a positive constant.
It is clear that for all values of $\lambda\, \geq\, 0$ the expansion
parameter $a$ obeys the inequality $0 \leq a<1$.
The variational parameter $C$ is fixed, for example, by requiring the
correct properties of the $\beta$--function in the infrared region,
as is discussed in more detail below.

The result of the expansion for the expression (17) to order $a^3$ is
\begin{eqnarray}
\label{Eq.20}
V\,&=&\,1\,+\,a\,A_3\,+
\,a^2\Bigl[\frac{1}{2}\,{A_3}^2\,+\,A_4\,+\,\frac{3}{2}A_3\,\Bigr]
\nonumber\\
&+&\,a^3\,\Bigl[\frac{1}{6} {A_3}^3\,+\,\frac{3}{2} {A_3}^2\,+
\,A_3 A_4\,+\,3 A_4\,+\,\frac{15}{8} A_3 \Bigr]\,+\,O(a^4)\, ,
\end{eqnarray}
where  $A_3\,=\,({\rm i}S_3)/P\, ,\,\,A_4\,=\,({\rm i}S_4)/P^2$ and
$P\,=\,C/(4\pi )^2$.

Within this approach the one--loop $\beta$--function is
\begin{equation}
\label{Eq.21}
{\beta}^{(3)} ({\lambda})\,=\,-\,\frac{b_0}{C^2}\,
\frac{a^4}{(2+a)\,(1-a)^2}\,(2+9\,a)\, .
\end{equation}
Here we must make some comments. The one--loop level that we use here
allows us to calculate the renormalization constant $Z_\lambda$ both
to $O(a^2)$ and to $O(a^3)$.
The expression (21) gives the $O(a^3)$ result, while
${\beta}^{(2)} ({\lambda})$ is obtained by simply omitting
the term $9\,a$ in the final bracket of Eq.~(21).
We will apply the phenomenon of induced convergence whereby
the variational parameter is not fixed for all orders but rather
depends on the order of the expansion.
As was empirically noticed in Ref.~[19] the results seem to converge if one
changes the variational parameter from order to order using some variational
principle. The induced convergence phenomenon has been discussed
in detail in Ref.~[20]. In Ref.~[14] the convergence of the optimized
$\delta$--expansion has been proved in the cases of zero and one dimensions.

The constant $C$ is determined by insisting that at an appropriately
large value of $\lambda$ the $\beta$--function should behave like
$-\,\beta(\lambda )/\lambda\,\simeq\,1$, which ensures the
correct infrared behaviour of the running coupling constant:
$\alpha_S(Q^2)\,\sim\, Q^{-2}$. This singular infrared behaviour of
the coupling constant leads to the linear part of the quark--antiquark
potential which is in agreement with the phenomenology of hadron spectroscopy.
In $O(a^2)$ and $O(a^3)$ we find
$C^{(2)}\,=\,0.977$ and $C^{(3)}\,=\,4.1$ respectively [21].

By using
the renormalization group equation
\begin{equation}
\label{Eq.22}
\ln \frac{Q^2}{Q_0^2}\,=\,\int _{{\lambda}_0}^{\lambda}\,
\frac{  d\,{\lambda}'  }{  {\beta} ({\lambda}')   }\, ,
\end{equation}
we obtain
\begin{equation}
\label{Eq.23}
\ln \,\frac{Q^2}{Q_0^2}\,\,=\,\frac{C^{(i)}}{2\,b_0}\, \Bigl[\,
f^{(i)}(a)\,-\,f^{(i)}(a_0)\, \Bigr]\, \, ,
\end{equation}
where
\begin{eqnarray}
\label{Eq.24}
 f^{(2)}(a) & = &  \frac{2}{a^2}\,+\,\frac{12}{a} \,+\,21\,\ln{
\frac{1-a}{a}}\,-\,\frac{9}{1-a} \, ,\nonumber \\
f^{(3)}(a) & = & \frac{2}{a^2} -   \frac{6}{a} - 48\ln a-
\frac{18}{11}\,\frac{1}{1-a} +  \frac{624}{121}\,\ln{(1-a) } +
\frac{5184}{121}\,\ln{( 1+\frac{9}{2}\,a )} \, .
\end{eqnarray}
The parameter $a_0$ and the momentum $Q_0$ in Eq.~(23) are defined by some
normalization point for our effective coupling constant
\begin{equation}
\label{Eq.25}
{\lambda}_t^{(3)}(q^2) \,=\,\frac{a^2}{C^{(3)}}\,(\,1\,+\,3\,a\,)\, .
\end{equation}
The effective coupling constant ${\lambda}_t^{(2)}(q^2) $ is obtained from
Eq.~(25) by omitting the term $3a$ in the final bracket and using the constant
$C^{(2)}$ instead of $C^{(3)}$.
To fix $a_0$ we use the value of the running coupling
constant measured at the $\tau$--lepton mass [2]:
$Q_0\,=\, M_{\tau}\,=\,1.777$ GeV and $\alpha_S(M_{\tau})\,=\,0.35$.
In Figure~2 we show the behaviour of our running expansion parameter
$a(Q^2)$ in the $t$--channel, obtained from (23) using $f^{(3)}(a)$.
As one can see, this parameter is indeed small in a wide range of $Q^2$,
and approaches 1 for very small $Q^2$.
Moreover,  the effective expansion parameter in the method is actually
$a^2(Q^2)$, with the $O(a^4)$ coming in at two--loop level.
Thus, the real expansion parameter is more or less
that shown in Figure~2.

Note that at large $Q^2\,=\,-\,q^2$ the effective coupling constant (25)
essentially coincides with the perturbative coupling constant. However, at
small $Q^2$ their behaviours are quite different. While the perturbative
coupling constant has a singularity at $Q^2\,=\,{\Lambda}^2$, the effective
coupling (25) is finite at all values of $Q^2\,\geq\,0$. As  follows
from Eq.~(23), the effective coupling constant (25) has a cut
along the positive $q^2$--axis. Thus, our approximation for the effective
coupling constant in the $t$--channel does not contradict the dispersion
relation (7), and we can make use of Eq.~(8) to find the effective coupling
constant in the $s$--channel.

By using Eq.~(8) one finds
\begin{equation}
\label{Eq.26}
\lambda_s^{(i)}(s)\,
=\,\frac{1}{2\pi{\rm i}}\,\frac{1}{2\,b_0}\,\bigl[\,{\phi}^{(i)}(a_{+})\,
-\,{\phi}^{(i)}(a_{-})\,\bigr]\, ,
\end{equation}
where
\begin{equation}
\label{Eq.27}
{\phi}^{(2)}(a)\,
=\,\frac{1}{1\,-\,a}\,\bigl[\,2\,-\,11\,a\,-\,4\,(1-a)\,\ln \, a\,+\,
3\,(1-a)\,\ln \, (1-a)\,\bigr]
\end{equation}
and
\begin{equation}
\label{Eq.28}
{\phi}^{(3)}(a)\,
=\,-4\,\ln \,a\,-\,\frac{72}{11}\,\frac{1}{1\,-\,a}\,+\,
\frac{318}{121}\, \ln\,(1\,-\,a)\,+\,\frac{256}{363}\,
\ln \,(1\,+\,\frac{9}{2}\,a)\,\, .
\end{equation}
The values of $a_{\pm}$ obey the following equation (from Eq.~(23))
\begin{equation}
\label{Eq.29}
f(a_{\pm})\,=\,f(a_0)\,+\,\frac{2\,b_0}{C}\,
\bigl[\,\ln\, \frac{s}{Q_0^2}\,\pm\,{\rm i} \,\pi\,\bigr]\,\, .
\end{equation}

In order to check our calculations we substituted $\lambda_s(s)$ obtained
from (26) into Eq.~(7), and found good agreement with the initial
$\lambda_t^{\rm eff}(q^2)$.
In Figure~3 we show the behaviour of ${\rm Re}\,a(s)$ and
$\,-\,{\rm Im}\,a(s)$ in the $s$--channel, working to
$O(a^3)$. For $O(a^2)$ we obtained
a similar result.
In Figure~4 we show the behaviour of the ratio
$R(s)\,=\,\lambda_s^{(i)}(Q^2)/\lambda_t^{(i)}(q^2)$ with
$s\,=\,Q^2\,=\,-\,q^2$ versus $s$ for $i\,=\,2$ (dashed line)
and $i\,=\,3$ (solid line). The value of $R(s)$ for $s\,=\, M_{\tau}^2$
is about 0.93, which represents a smaller correction than the perturbative
$\pi^2$-correction of Eq.~(10).
The agreement between successive orders is confirmation of the mechanism
of induced convergence. As follows from the results we have obtained, the
effects of analytic continuation can be essential for
the definition of the QCD coupling constant from timelike processes,
for instance from $\tau$--decay. We plan to discuss this problem in future
publications.


\vspace{1.0cm}
The authors would like to thank D.I.~Kazakov, A.N.~Sissakian and
O.P.~Solovtsova for interest in the work and useful comments.
One of us (H.F.J.) is grateful to the Royal Society for a travel grant,
the other (I.L.S.) would like to thank  the Russian Science Foundation
(~Grant~93-02-3754~) for financial support.

\newpage
\vspace{1.0cm}
\begin{center}
{\bf Figure Captions}
\end{center}

\begin{tabbing}
\vspace{1.0cm}
{\bf Figure 1 }
\=The integration contour in Eq.~(4).\\
\end{tabbing}

\begin{tabbing}
\vspace{1.0cm}
{\bf Figure 2 }
\=The behaviour of our running expansion parameter
$a(Q^2)$ in the $t$--channel.\\
\end{tabbing}

\begin{tabbing}
\vspace{1.0cm}
{\bf Figure 3 }
\=The behaviour of ${\rm Re}\,a(s)$ and $\,-\,{\rm Im}\,a(s)$
in the $s$--channel at $\O(a^3)$.\\
\end{tabbing}

\begin{tabbing}
\vspace{1.0cm}
{\bf Figure 4}
\=The behaviour of the ratio
$R(s)\,=\,\lambda_s^{(i)}(Q^2)/\lambda_t^{(i)}(q^2)$ with
$s\,=\,Q^2\,=\,-\,q^2$\\
 \>versus $s$ for $i\,=\,2$ (~dashed line~)
and $i\,=\,3$ (~solid line~).
\end{tabbing}

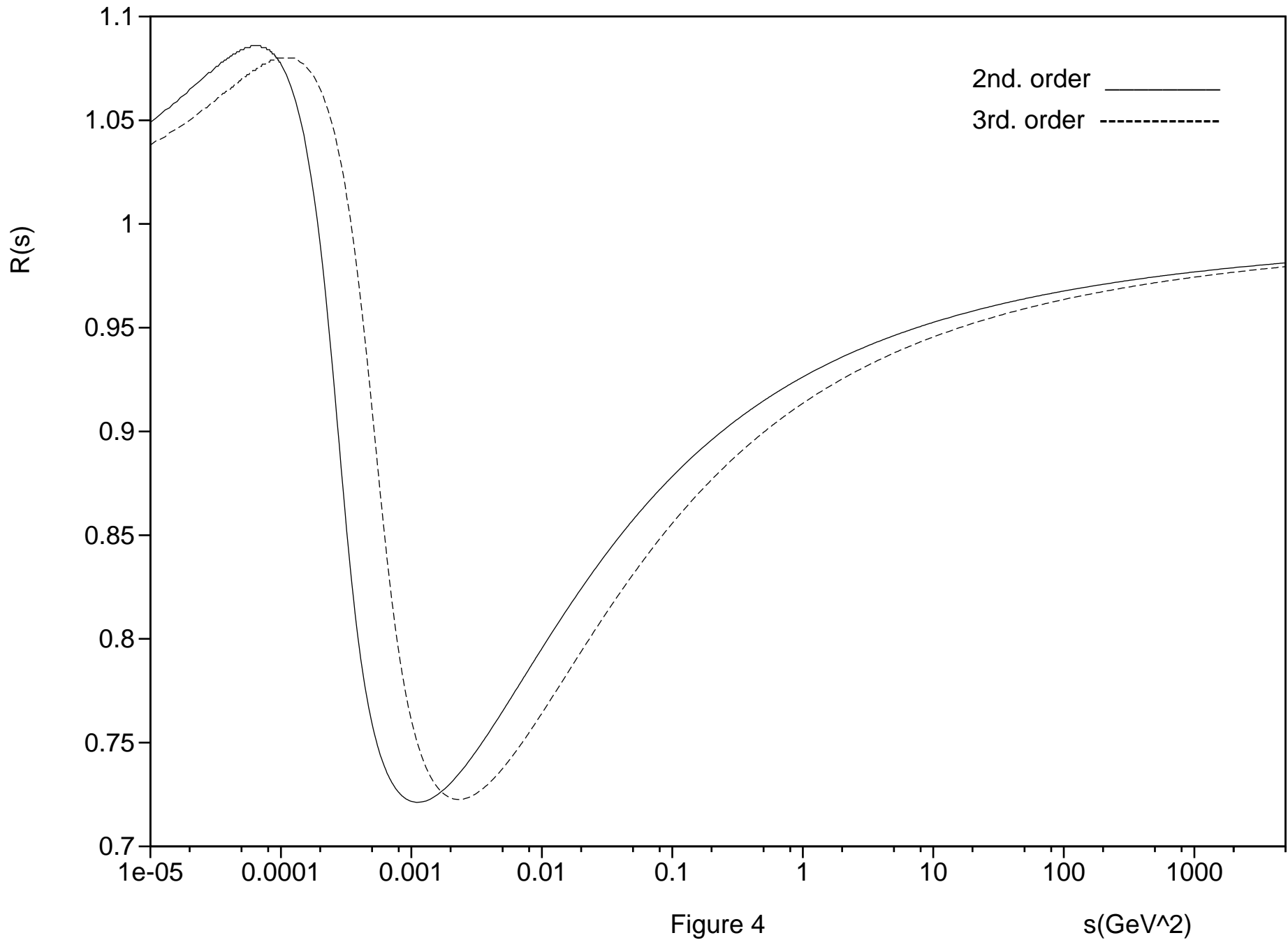
\end{document}